\documentclass[11pt,a4paper]{article}   

\usepackage{amsmath}    
\usepackage{amsfonts}  
\usepackage{amssymb}
\usepackage{bbm,dsfont}
\usepackage{graphicx}   
\usepackage[T1]{fontenc}

\usepackage{float}   
\usepackage{bm}
\usepackage{hyperref}

\usepackage{color}   
\usepackage{url}
\usepackage{soul}    
\usepackage[a4paper]{geometry} 
\graphicspath{{./images/}}

\definecolor{lightblue}{rgb}{0.8,0.9,1} 

\begin{document}

\title{An experiment to improve understanding wave-particle duality}
\author{Michel Gondran$^1$ and Alexandre Gondran$^2$\\
\small$^1$ Fondation Louis de Broglie, Paris, France\\
\small\texttt{michel.gondran@polytechnique.org}\\
\small$^2$ ENAC, Universit\'e de Toulouse, Toulouse, France\\
\small\texttt{alexandre.gondran@recherche.enac.fr}}


\maketitle
\begin{abstract}

This article presents an experiment that can be conducted today and that could provide a deeper understanding of the interaction between the wave and particle aspects of an atom. The wave-particle duality is often presented as mutually exclusive: one considers either the wave aspect or the particle aspect. Our proposed experiment involves both aspects simultaneously and raises new questions. It is a slightly modified version of Young’s double-slit interference experiment (a grid of narrow slits is added between the two wide slits) and is carried out using Rydberg atoms. Young-type interference experiments typically involve only the de Broglie wave $\psi$, which depends solely on the mass and velocity of the atoms.
However, with Rydberg atoms having a large principal quantum number, the “size” of the atom-particle also becomes significant. The two large slits are wide enough to allow the Rydberg atoms to pass through, whereas the grid of narrow slits prevents them from passing through.

We numerically simulate the possible outcomes based on different hypotheses regarding wave-particle interaction. Conducting the experiment in practice would allow us to distinguish between these hypotheses and deepen our understanding of wave-particle interaction. 

The conceptual framework of Louis de Broglie’s double solution theory is well-suited to this experiment because it distinguishes between two types of waves: an external or statistical wave (de Broglie’s wave) and an internal or physical wave (corresponding to the physical particle). We will examine the relevance of this approach.





\end{abstract}

\section{Introduction}
Wave-particle duality is one of the defining characteristics of quantum mechanics. It is clearly demonstrated in Young's double-slit  experiment performed with massive particles: electrons, atoms, or molecules are emitted one by one from a coherent source and then pass through a plate with two slits (or holes); on a screen placed further away, we observe the impacts (small scintillations) of the particles arriving one by one. The cumulative sum of the impacts produces interference fringes that can only be explained by a wave phenomenon.  
Below we recall John Bell’s commentary in his article ``Six possible worlds of quantum mechanics''~\cite{Bell1992} and the causal interpretation given by Louis de Broglie~:
\begin{quote}
    ``While the founding fathers agonized over the question
    \textbf{"particle" or "wave"},
    de Broglie in 1925 proposed the obvious answer
    \textbf{"particle" and "wave"}.
It is not clear from the smallness of the scintillation on the screen that we have to do with a particle~? And is it not clear, from the diffraction and interference patterns, that the motion of the particle is directed by a wave~? De Broglie showed in detail how the motion of a particle, passing through just one of two holes in a screen, could be influenced by waves propagating through both holes. And so influenced that the particle does not go where the waves cancel out, but is attracted to where they cooperate. This idea seems to me so natural and simple, to resolve the wave-Particle dilemma in such a clear and ordinary way, that it is a great mystery to me that it was so generally ignored. Of the founding fathers, only Einstein thought that de Broglie was on the right lines. Discouraged, de Broglie abandoned his picture for many years. He took it up again only when it was rediscovered, and more systematically presented, in 1952, by David Bohm. [...] 

The de Broglie Bohm synthesis, of particle and wave, could be regarded as a precise illustration of Bohr's complementarity ... if Bohr had been using this word in the ordinary way. This picture combines quite naturally both the waviness of electron diffraction and interference patterns, and the smallness of individual scintillations, or more generally the definite nature of large-scale happenings.''~\cite{Bell1992}
\end{quote}

Louis de Broglie proposed~\cite{deBroglie1927} in 1927 to complete the formalism by adding a second wave to the usual \textbf{\emph{statistical wave}}, $\bm{\psi}$, solution of the Schrödinger equation.
This second wave, $\bm{u}$, which he calls the \emph{matter wave} or \textbf{\emph{physical wave}}, corresponds to the mass density of the extended corpuscle. Unlike the statistical wave, which can spread throughout all space over time, the matter wave $u$ has a finite size and remains localized. Furthermore the center of gravity $\textbf{X}(t)$ of $u$-wave is guided by the statistical $\psi$-wave. These two waves, $\psi$ and $u$, are in resonance due to a phase match.
This theory, known as the \emph{double-solution theory}~\cite{deBroglie1956,deBroglie1987,Fargues2017,Colin2017,Gondran2023d} is an ingenious way of completing the formalism, making it causal, without contradicting experimental measurements~\footnote{It should be specified that Schr\"odinger's statistical $\psi$-wave was usually referred to in the 20th century as the de Broglie matter wave. However, for de Broglie, the matter wave that he defined in 1927 is not the statistical wave but the $u$-wave corresponding to the spatial extension of the particle that concentrates its mass. The statistical $\psi$-wave was called the \textit{pilot wave} by de Broglie in 1927. In the remainder of this article, we prefer to use the term \textit{physical wave} (or \textit{internal wave}) rather than \textit{matter wave} to avoid confusion.}. 

If we reduce the physical $u$-wave to its center of gravity $X(t)$, we obtain the \emph{pilot wave theory}, presented by de Broglie, at the Solvey Congress in 1927, which for him was only a second-best solution. His work on the pilot wave was continued from 1952 onwards, notably by David Bohm~\cite{Bohm1952}, under the name de Broglie-Bohm (dBB) theory. Figure~\ref{fig:3theories} sums up these different approaches to quantum mechanics.

\begin{figure}[h!]
    \centering
\fbox{%
\begin{minipage}{0.7\textwidth}
\begin{center}
   Double solution theory~: $\psi(\textbf{x},t) $ and an extended particle $ u(\textbf{x},t)$\\
   \fbox{%
   \begin{minipage}{0.9\textwidth}
   \begin{center}
   De Broglie-Bohm theory~: $\psi(\textbf{x},t)$ and point particle $\textbf{X}(t)$\\
   \fbox{%
   \begin{minipage}{0.8\textwidth}
   \begin{center}
    Orthodox theory~: $\psi(\textbf{x},t)$ only
    \end{center}
   \end{minipage}
   }
\end{center}
\end{minipage}
}
\end{center}
\end{minipage}
}
    \caption{The de Broglie-Bohm (dBB) theory adds the position $\textbf{X}(t)$ of the center-of-mass to the orthodox description of a particle by its statistical wave function $\psi(\textbf{x},t)$~ \cite{deBroglie1927}. The double solution theory adds to the dBB description a physical wave function, $u(\textbf{x},t)$, for which the center of mass is the dBB position $\textbf{X}(t)$.}
    \label{fig:3theories}
\end{figure}

The advantage of defining an additional internal wave in the double-solution theory is that it explicitly defines a spatial extension $u$ of the particle, corresponding to its physical size. This hypothetical $u$ can be seen as a local hidden variable for the standard theory. The interference experiment we propose in this article explicitly takes this hypothetical ``size'' into account. The results of the experiment will differ depending on whether or not this hypothetical size exists. 
However, the theory of the double-solution was never fully completed, and Louis de Broglie does not specify whether the total wave function should be the product or the sum of $\psi$ and $u$, or simply the pair $(\psi, u)$, or even something else, as Albert Einstein pointed out in a letter dated 1954 in which he supports de Broglie’s approach~\cite{Einstein1953}~\footnote{\textit{``The point of view you take in your note seems very clear. You don't believe, if I understand correctly, in the possibility of adopting the program put forward again by Mr.~Bohm:
    a) Solution of the Schrödinger equation for a $\psi$ field
    b) Addition of a trajectory compatible with the $\psi$ function.
Instead, you propose a representation of physical reality (complete description) of the form $\Psi= \psi \cdot u$.
This constitutes a form of product in which one of the factors leads to the particle structure and the other to the wave structure. This would in fact be a satisfactory representation of the dual structure imposed on us by experience. It would be a truly new theory, not a complement to the old ones.''}~\cite{Einstein1953}}.

\begin{figure}[ht]
\begin{center}
\includegraphics[width=0.49\linewidth]{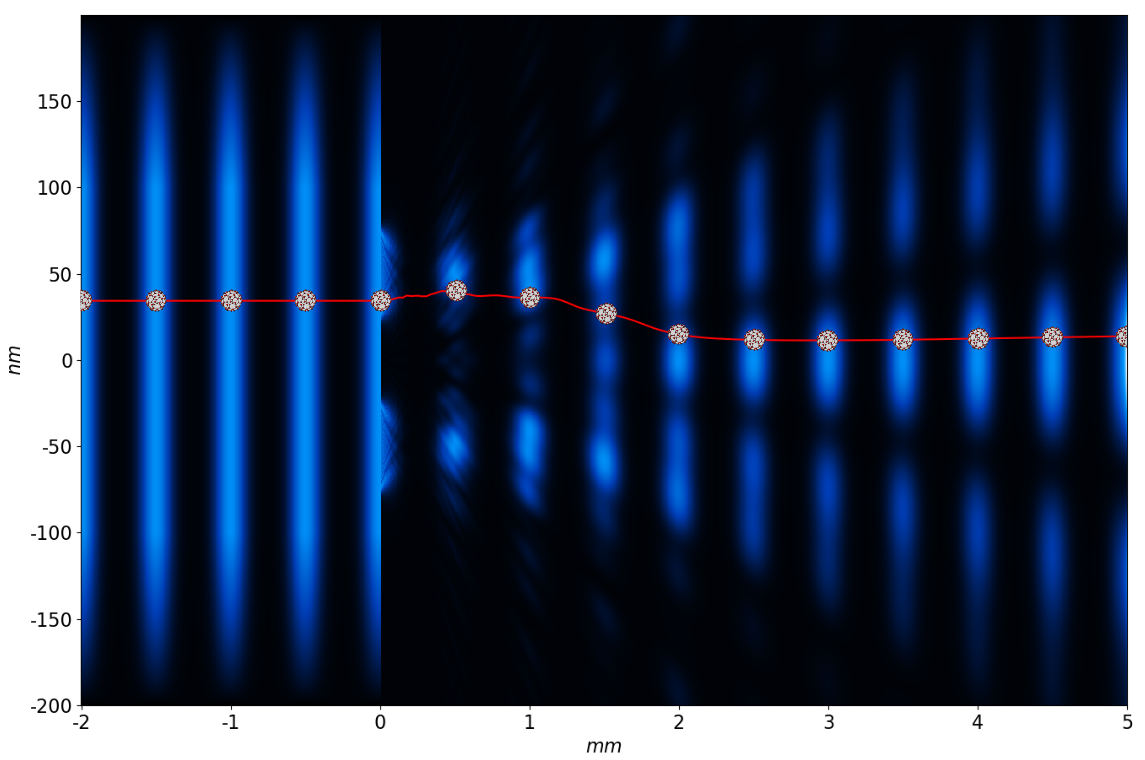}
\includegraphics[width=0.49\linewidth]{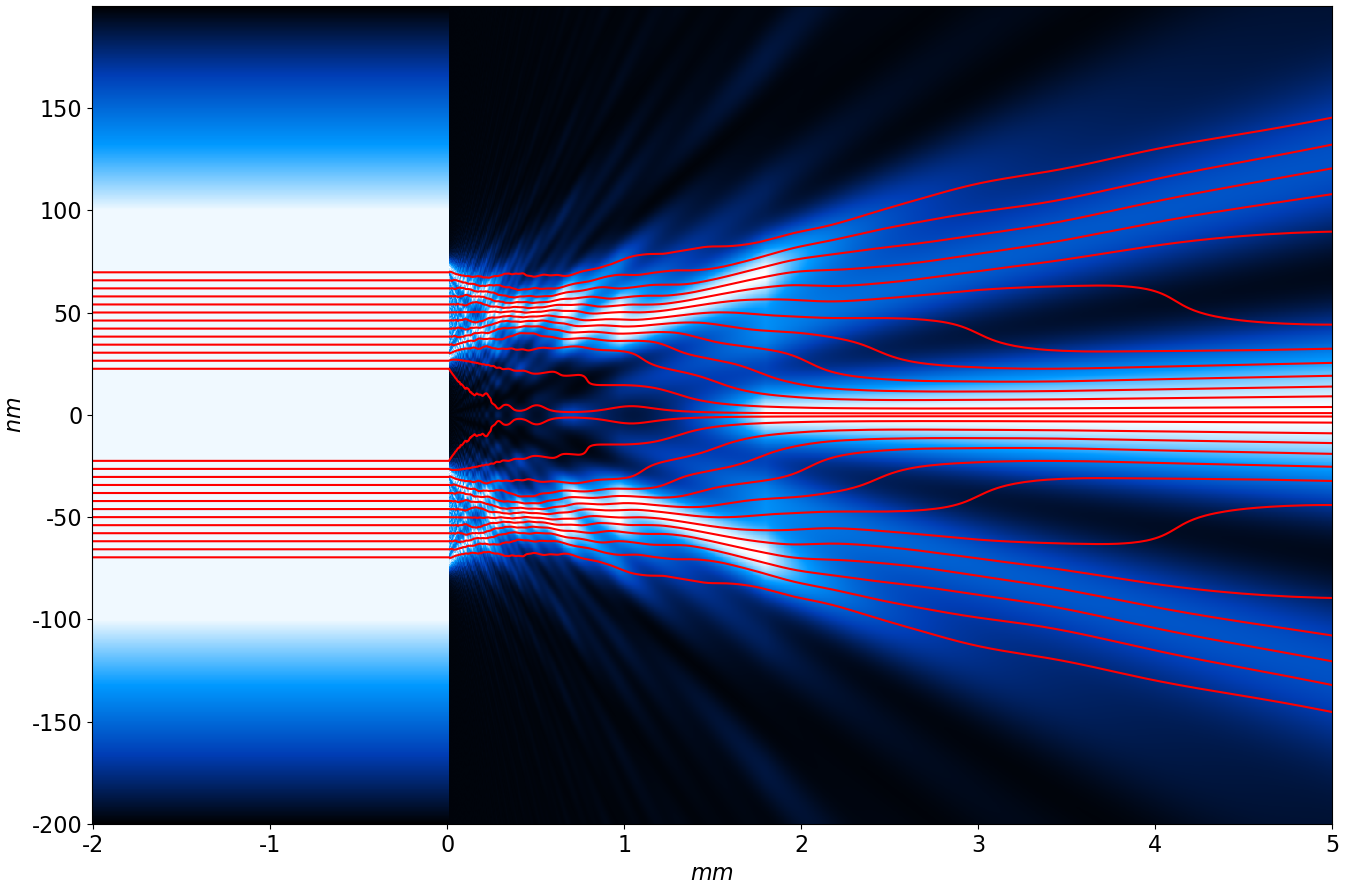}
\caption{\label{fig:dBBtrajectories}\textbf{Left:} Evolution of the probability density $|\psi(x,y,t|^2$ of a $C_{60}$ molecule for Young's double-slit experiment. Simulations performed under the conditions of the experiment conducted by Zeilinger's team~\cite{Nairz2003}. The statistical wave density $\psi(x,y)$ of the molecule evolves from left to right. Fifteen densities are shown at various times (every $\Delta t=2.5\mu s$ i.e. every $0.5mm$ on x-axis) before the slits (left) and after the slits (right). The two slits are located at $0mm$ on the x-axis and are spaced $100nm$ on the y-axis. The brighter the blue, the higher the density. The trajectory of the center of mass of a $C_{60}$ molecule is shown in red; its initial position is randomly selected within the density. The $C_{60}$ molecule is not a point particle; it has a size corresponding to its physical $u$-wave (in the figure, its size is  magnified 13 times). According to the double-solution theory, the $u$-wave is guided by the $\psi$-wave.\newline\textbf{Right: } Another representation of the same density $|\psi(x,y,t|^2$ as in the figure on the left. Here, the evolution is shown at every time step $t=x/v_x$; in red, 24 dBB trajectories of $C_{60}$ corresponding to 24 different starting points of the center of mass of a $C_{60}$. }
\end{center}
\end{figure}

The trajectories of particles in the dBB theory~\cite {Bohm1952,Holland1993,Goldstein2009} have been simulated for photons~\cite{Ghose2001,Gondran2010a,Foo2022,Sanz2012}, electrons~\cite{Philippidis1979,Gondran2001,Sanz2014,Gondran2014d}, atoms~\cite{Gondran2005a}, or molecules~\cite{Gondran2021}. Figures~\ref{fig:dBBtrajectories} show the trajectories of $C_{60}$ molecules for double-slit experiment performed by Zeilinger's team~\cite{Nairz2003}. The molecule is guided by the statistical wave $\psi$ (in blue), a solution to the time-dependent Schrödinger equation\footnote{Animations have been created for interference with $C_{60}$ molecules~: \url{https://vimeo.com/350132498}}.

In this article, we adopt the framework of Louis de Broglie’s double-solution theory and propose an experiment that clarifies the interaction between the wave and the particle in this approach.

In matter interference experiments, the statistical $\psi$-wave alone is sufficient to explain the interference patterns, as Hornberger et al.~\cite{Hornberger2013} points out; they demonstrated this phenomenon using very large $C_{60}F_{48}$ molecules~: 
\begin{quote}
 ``We have seen that the de Broglie wave [i.e. the statistical $\psi$-wave] well describes the center-of-mass motion of even very complex particles, giving rise to interference phenomena which can be surprisingly robust against a large variety of internal state transformations and against interactions between external force fields and internal particle dynamics [i.e. the physical $u$-wave].''~\cite{Hornberger2013}
\end{quote}
The ``size'' of the particles is not taken into account or is only implicitly considered: the width of the slits must be much greater than the size of the particles. 
For example, in the double-slit experiment conducted by Arnt et al.~\cite{Arndt1999} using $C_{60}$ fullerene molecules: the width of the slits is 55 times greater than the size of the $C_{60}$ buckyball, i.e. approximately 1~nm. This size, 350 times greater than its de Broglie wavelength, is at least 500 times smaller than the width of the \textit{statistical $\psi$-wave} function $\psi$ that passes through the two slits.    
For double-slit experiments performed with molecules, atoms, neutrons or electrons, the physical size~\footnote{By physical size, we mean the size of the support of the $u$-wave.} is generally not taken into account. In the experiment by Schimizu et al.~\cite{Shimizu1992} with ultra-cold silver atoms, the size of the atoms is approximately $2a_0\sim 1$\AA (with  $a_0=5.29\times 10^{-11}m$ the Bohr radius) while the slits are 20,000 times wider.
More recently, matter-wave interferometry experiments with Rydberg atoms have been performed using an electric Rydberg-atom interferometer~\cite{Palmer2019a,Palmer2019b,Tommey2021} rather than a grating. We propose a grating-based interference experiment since we need to explicitly account for the physical size of atoms.

On the other hand, the experiments conducted by Fabre et al.~\cite{Fabre1983} involve \textit{``measuring atomic dimensions by transmission of Rydberg atoms through micrometre size slits''}.   
The transmission through a grating of slits (each $2 \mu m$ wide) became zero when the principal quantum number $n\geq 60$. They showed that Rydberg atoms behave like \textit{``hard spheres''} of diameter $d=k a_0 n^2$, with $k=9$ experimentally fixed. 
The Rydberg atoms are destroyed when they come into contact with the metal edge of a slit. They are ionised during the slit crossing. 

Our experiment is a variation on the double-slit experiment performed with Rydberg sodium atoms, in which a grating of very narrow slits is added between the two slits. 
The idea of using asymmetric slits was first proposed in Bo{\v z}i{\'c} et al.~\cite{Bozic2004}.
As in~\cite{Fabre1983}, the narrow slits in the grating are small enough to prevent Rydberg atoms from passing through. The interference patterns would differ depending on the assumption we make regarding wave-particle duality.
Within the framework of the double-solution theory, our experiment deepens our understanding of the interaction between the statistical $\psi$-wave  and the physical $u$-wave of an atom; we define three possible hypotheses regarding this interaction. These three hypotheses are mutually exclusive:
\begin{itemize}
    \item[]\textbf{Hypothesis 1:} 
    The physical $u$-wave can be neglected, ignored, or treated as point-like, and only the statistical $\psi$-wave needs to be considered in the interference experiments. This hypothesis does not seem realistic in light of the experiment~\cite{Fabre1983} but we present it for reference. 
    \item[]\textbf{Hypothesis 2} The physical $u$-wave and the statistical $\psi$-wave are inextricably linked. If the $u$-wave cannot pass through a slit, then the $\psi$-wave is also stopped by the slit.
    In dBB theory, there is an action of the statistical wave $\psi$ on the particle's center-of-mass $X(t)$ (the $\psi$-wave is called the pilot wave) and thus on the $u$-wave. This hypothesis further assumes that there is a feedback from the physical wave $u$ to the statistical wave $\psi$. If the $u$-wave is stopped, then the $\psi$-wave is also stopped. There is no \emph{empty wave} in this hypothesis~\cite{Selleri1989}; it is like if the slits of the grid does not exist. What ever the size of the slits, if there are too narrow to not let $u$-wave through, then nothing else gets through (at least not the $\psi$-wave).
    \item[]\textbf{Hypothesis 3} Even if the physical $u$-wave is unable to pass through a slit, the statistical $\psi$-wave passes through it and continues to propagate beyond; it can thus interfere with other parts of itself that have passed trough others slits.
\end{itemize}

There are at least three definitions of the \emph{\textbf{empty wave}} in dBB framework~: the first one is the most common~\cite{Hardy1992}. An empty wave is part of the statistical $\psi$-wave within which there is no particle (i.e. no $u$-wave); for example, in the case of the double-slit experiment with slits $A$ and $B$, after the slits, we have: $\psi=\psi_A+\psi_B$, where $\psi_A$ (and $\psi_B$, respectively) is the branch of the statistical $\psi$-wave that continues beyond slit $A$ (and beyond slit $B$, respectively). According to the pilot wave approach, if a first particle (i.e., the $u$-wave) passes through slit $A$, then $\psi_B$ is called the empty wave and $\psi_A$ the full (non-empty) wave. If a second particle passes through slit $B$, it is $\psi_A$ that is called the empty wave this time.
$\psi_A$ and $\psi_B$ are sometimes considered empty and sometimes 'full' depending of the position of the $u$-wave. After passing through the slits, the empty wave interferes with the full wave, producing the well-known interference pattern ($|\psi_A+\psi_B|^2$). 
The question of whether or not this empty wave exists does not arise: it is simply a definition within the framework of dBB theory.

A second definition is more disputable. An empty wave is the $\psi$-wave when the particle has disappeared; for example, if the particle has been stopped (collapsed) by the plate with two slits, the statistical $\psi$-wave that continues to propagate beyond the slits is called an empty wave or \textbf{\textit{ghost wave}} to distinguish it from the first definition. Indeed, in dBB approach, there is no collapse of the $\psi$-wave function, then the ghost $\psi$-wave can propagate indefinitely. Does this ghost wave can interfere with a other $\psi$-wave coming from a other particle~? Selleri~\cite{Selleri1989} reported proposal experiments supporting this view.

A third definition, which is a special case of the first definition, is provided by Hypothesis~3. The term \textbf{\textit{phantom empty wave}} is used to distinguish it from the first definition. 
A \textit{phantom empty wave} is a part of the statistical $\psi$-wave that will never be followed by a dBB trajectory. It is an empty wave in the sense of first definition, but it will always remain empty.

Return to the double-slit experiment involving slits $A$ and $B$. However, if slit $A$ is now so narrow that it cannot allow the $u$-wave (i.e. the particle) to pass through, then $\psi_A$ can never be a full (or non-empty) wave. On the other hand, when the particle passes through slit $B$, $\psi_A$ is an empty wave called \textit{phantom empty wave} because it will never be full (or always empty) regardless of the particle's position within the $\psi$-wave.
We can then legitimately question the existence of this phantom empty wave, and that is the purpose of the experiment proposed here. 
Indeed, after the slits, if the phantom empty wave, $\psi_A$, does not exist (Hypothesis 2), then everything happens as if slit $A$ did not exist, and we would therefore obtain on the detection screen only the density $|\psi_B|^2$ (diffraction by a single slit).

On the other hand, if the phantom empty wave, $\psi_A$, exists (Hypothesis 3), it should interfere with the full wave $\psi_B$ and thus alter the trajectories of the particles that passed through slit $B$. The interference pattern obtained on the detection screen will be the density $|\psi_A+\psi_B|^2$ truncated of the particle trajectories that should have passed through slit $A$ (since nothing passed through slit $A$). Indeed, if slit $A$ is 9 times smaller than slit $B$, then 10\% of the density $|\psi_A+\psi_B|^2$ must be truncated: the 10\% corresponding to the trajectories that should have passed through $A$'slit\footnote{Recall that when the particle source is coherent, the trajectories of dBB do not cross. This implies that each trajectory of dBB corresponds to a \emph{quantile trajectory} of the particle presence density (i.e. \emph{time-dependent quantile position})~\cite{Brandt1998,Oriols1996,Brandt2001,Coffey2008,Gondran2018}. In Young’s double-slit experiment, where slit $A$ is 9 times smaller than slit $B$, the decile  (10\% quantile) of the density on the detection screen divides the particle impacts into two groups: those that passed through slit $A$ (10\%) and those that passed through slit $B$ (90\%).}.



\begin{figure}[ht]
\begin{center}
\includegraphics[width=0.99\linewidth]{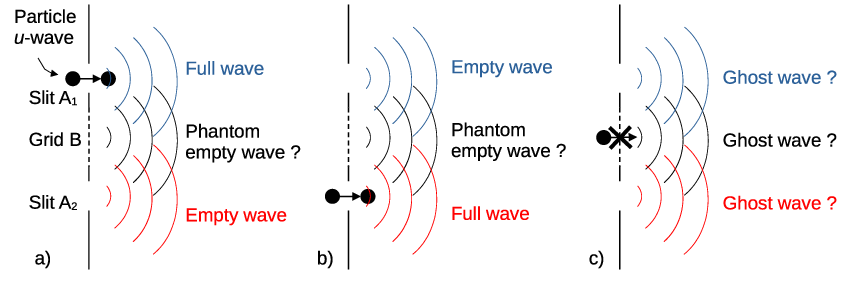}
\caption{\label{fig:empty_wave}Illustration of the three types of empty waves in the wave-pilot theory as applied to our experiment. The first two figures, a) and b), show the case where the particle (i.e. the $u$-wave) passes through one of the wide slits, $A_1$ or $A_2$. In the case of Figure a), where the particle passes through $A_1$, the wave $\psi_{A_1}$ (in blue) is the ``full'' wave and $\psi_{A_2}$ (in red) is the empty wave. $\psi_B$ (in black) is the phantom empty wave, and its existence is being investigated in our experiment. Figure b) is the mirror image of Figure a), with the particle passing through slit $A_2$. In Figure c), the $u$-wave (i.e. the particle) cannot pass through the slits of grid $B$; it is stopped by the plate, and all waves that continue beyond it are called ghost waves.}
\end{center}
\end{figure}

It is this third definition that we examine in this article. Figure~\ref{fig:empty_wave} illustrates the three types of empty waves for our experiment. Hypothesis~3 assumes the existence of phantom empty waves, unlike Hypothesis~2.
The experimental conditions described in the next section are chosen to shed as much light as possible on the differences in interference fringes.

\section{Proposed experiment: Interference of Rydberg atoms through a double-slit and an additional grating of narrow slits.}\label{section:diffraction}

\begin{figure}[ht]
\begin{center}
\includegraphics[width=0.6\linewidth]{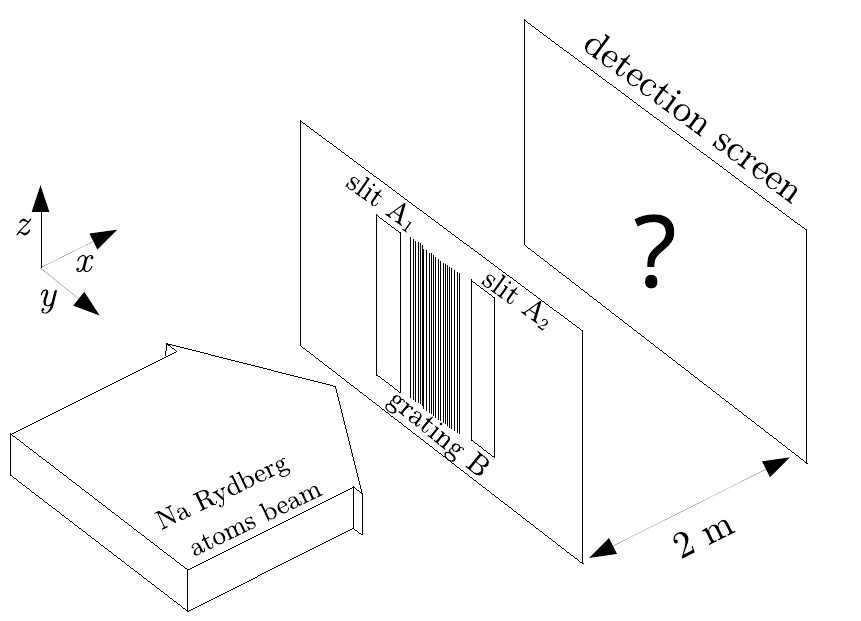}
\caption{\label{fig:schema} Schematic drawing of the interference experiment.}
\end{center}
\end{figure}

The proposed experiment corresponds to the interference of a beam of Rydberg atoms through slits of different sizes: two large identical slits $A_1$ and $A_2$, with a width equal to $4\mu m$ and centred at $\pm8\mu m$, and a grating $B$ of 40 narrow slits, with a width of $0.1\mu m$ and spaced $0.2\mu m$ apart (center-to-center), the grating being placed between the two slits, cf. Fig~\ref{fig:schema}.
The beam of Rydberg atoms are sodium atoms of mass $m=3.8\times 10^{-29}kg$ with a velocity $v_x=150m/s$ along the $(Ox)$ axis corresponding to a de Broglie wavelength $\lambda_{dB}=1.15\times 10^{-10}m$.

If the atoms are not excited into a Rydberg state, their \textit{sizes} are small (i.e. the size of the support of $u$-wave $\sim1$\AA) and they can pass trough the slits of the grating: the interference pattern will be different from that observed to the double-slit experiment alone. 
However, with Rydberg atoms, which cannot pass through the slits of the grating, we either observe the \emph{same interference pattern as in the double-slit experiment alone} (Hypothesis 2), or we observe a very different interference pattern with a \emph{dark band in the center} (Hypothesis 3). This can be easily explained with the framework of the double-solution theory.

This experiment builds on an earlier proposal~\cite{Gondran2007,Bozic2004}, 
but here we focus our analysis on the double-solution interpretation and wave-particle interaction. From another perspective, asymmetrical double-slit experiments for photons was proposed~\cite{VoVan2022}.


\subsection{Interference patterns of unexcited sodium atoms}
\begin{figure}[ht]
\begin{center}
\includegraphics[width=0.42\linewidth]{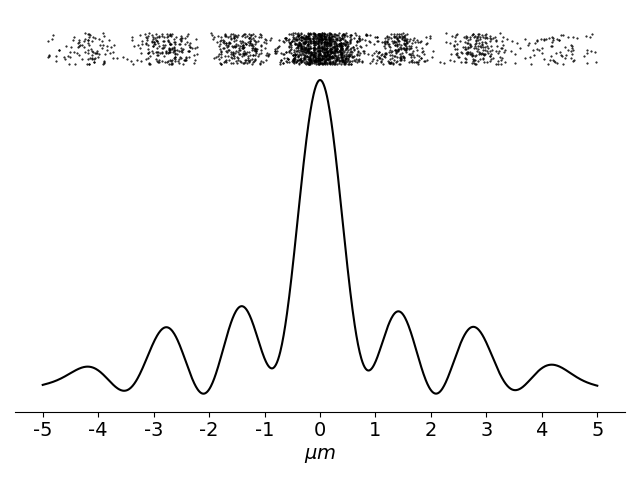}
\includegraphics[width=0.52\linewidth]{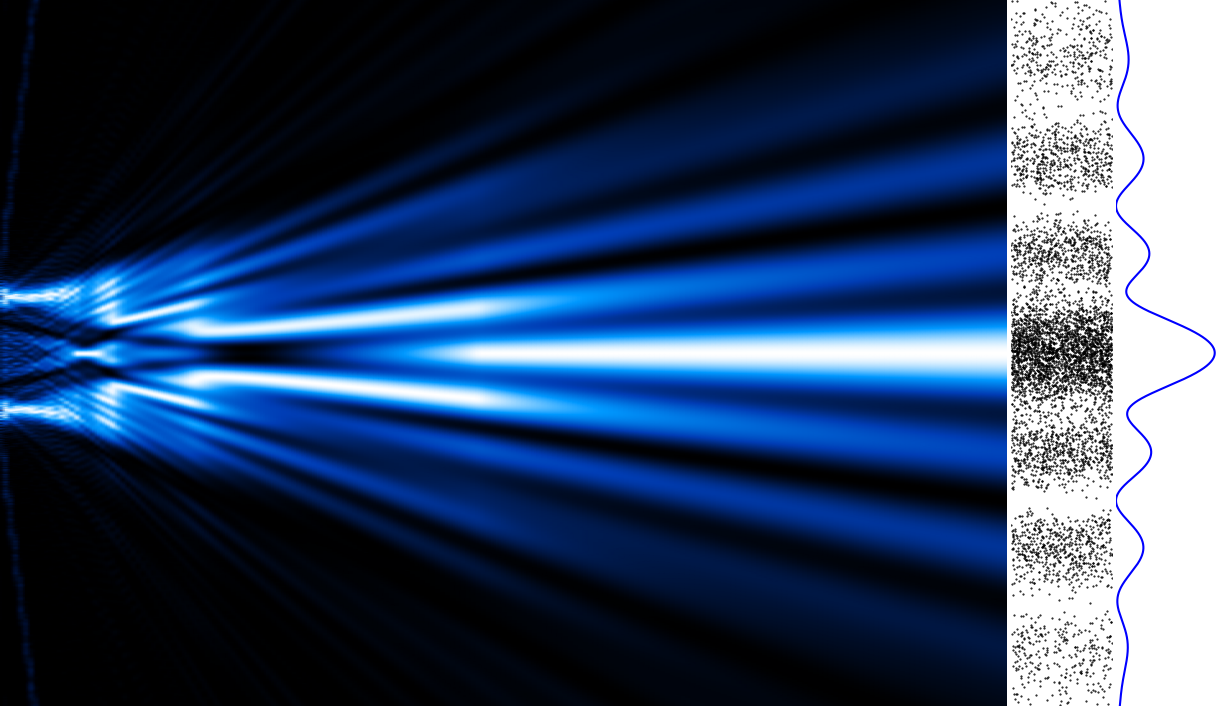}
\caption{\label{fig:results-unexcited} Simulation results for Hypothesis~1: identical to those obtained for unexcited sodium atoms. \textbf{Left:} The interference pattern corresponding to the norm of the statistical wave function, $|\psi(y)|^2$ of the unexcited sodium atoms on the detection screen at 2m after the slits. This pattern is common to all interpretations of quantum mechanics. Above the density curve, 5,000 atomic impacts show what is actually observed.
\textbf{Right:} 
Evolution of the probability density $|\psi(x,y,t=x/v_x)|^2$ of an unexcited sodium atom, from the slits to a distance of 2m beyond them. For $x=2m$ (on the right),  the  density shown in the figure on the left is obtained.}
\end{center}
\end{figure}

Before considering what happens with Rydberg atoms, let us see what happens if the experiment is carried out with unexcited sodium atoms passing through the narrow slits of $B$ grating and the double-slit $A_1-A_2$. Regardless of interpretation of quantum mechanics, two meters after the slits, we would obtain the interference pattern (i.e. the norm of the statistical wave function, $|\psi(y)|^2$) shown in Figures~\ref{fig:results-unexcited}: a large central peak and six much smaller ones (figure on the left); on the right the evolution of $|\psi(x,y,t)|^2$ at all distance after the slits (with $x=v_x\times t$). 

Calculations of the time evolution of the statistical $\psi$-wave are performed by numerically solving the time-dependent Schr{\"o}dinger equation~\cite{Gondran2005a,Philippidis1979} using Feynman path integrals~\cite{Feynman1965,Darwin1927}. 
This statistical $\psi$-wave initially from the sodium source is large enough to pass through all the slits at the same time $t_0$ (the two large as well as the 40 small ones). After the slits, $\psi$-wave divides, then recombines and forms interference patterns. Feynman path integrals give the value of $\psi$ at a distance of $x$ meter beyond the slits (at $t=x/v_x$) and along the $y$-axis (perpendicular to the slits): 
\begin{equation}
\psi(y,t)=\int_{A_1\cup A_2\cup B}K(y,t\,;\,y_0,t_0)\,\psi(y_0,t_0)\,dy_0   
\end{equation}
where $K$ is the kernel and $\psi(y_0,t_0)$ is the wave function at the slits (which can be considered constant and equal to 1). The integration is done on points $y_0\in A_1\cup A_2\cup B$  (on slits $A_1$, $A_2$ and the slits of $B$ grid). $\psi(y,t)=\psi_{A_1}(y,t)+\psi_{A_2}(y,t)+\psi_B(y,t)$ if we note for $i\in\{A_1, A_2, B\}$: $\psi_i(y,t)=\int_{i}K(y,t\,;\,y_0,t_0)\,\psi(y_0,t_0)\,dy_0$. 
The kernel between two events $a=(y_a, t_a)$ and $b=(y_b, t_b)$ is~: $K(b; a)\,\sim \,\exp\left({\frac{i}{\hbar}S_{cl}(b;a)}\right)$
with $\int_{-\infty}^{+\infty}K(b; a)\,dy_a=1$, so we have~:
$$
K(b; a) = \left({\frac{2\pi i\hbar(t_b-t_a)}{m}}\right)^{-1/2} \exp{\left(\frac{im(y_b-y_a)^2}{2\hbar(t_b-t_a)}\right)}.
$$
Indeed, the classical action $S_{cl}$ for a free particle is~:
$S_{cl}(b; a)=\int_{t_a}^{t_b}L(\dot{y},y,t)\,dt=\frac{m(y_b-y_a)^2}{2(t_b-t_a)}$
where $L$ is the lagrangien of a free particle: $L(\dot{y},y,t)=\left(m\,\frac{{\dot{y}}^2}{2}\right)$.

\subsection{Hypothesis 2: Results identical to those for the interference patterns with only double-slit $A_1$ and $A_2$}
We now consider Rydberg atoms with physical \emph{size} around $0.1\mu m$, corresponding to a principal quantum number $n=15$ (we use the equation $d = ka_0n^2$, with $k = 9$~ from~\cite{Fabre1983}). These Rydberg atoms are chosen so that they cannot pass though the $B$ grating.
According to Hypothesis~1, in which the Rydberg atoms are treated as point-like, we obtain the densities shown in Figures~\ref{fig:results-unexcited}.

\begin{figure}[ht]
\begin{center}
\includegraphics[width=0.42\linewidth]{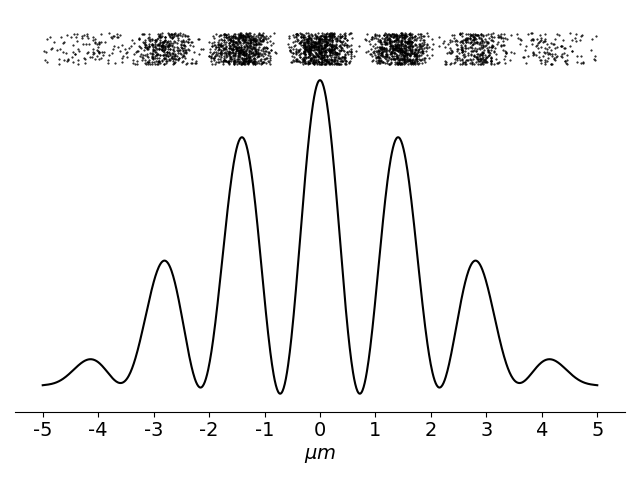}
\includegraphics[width=0.52\linewidth]{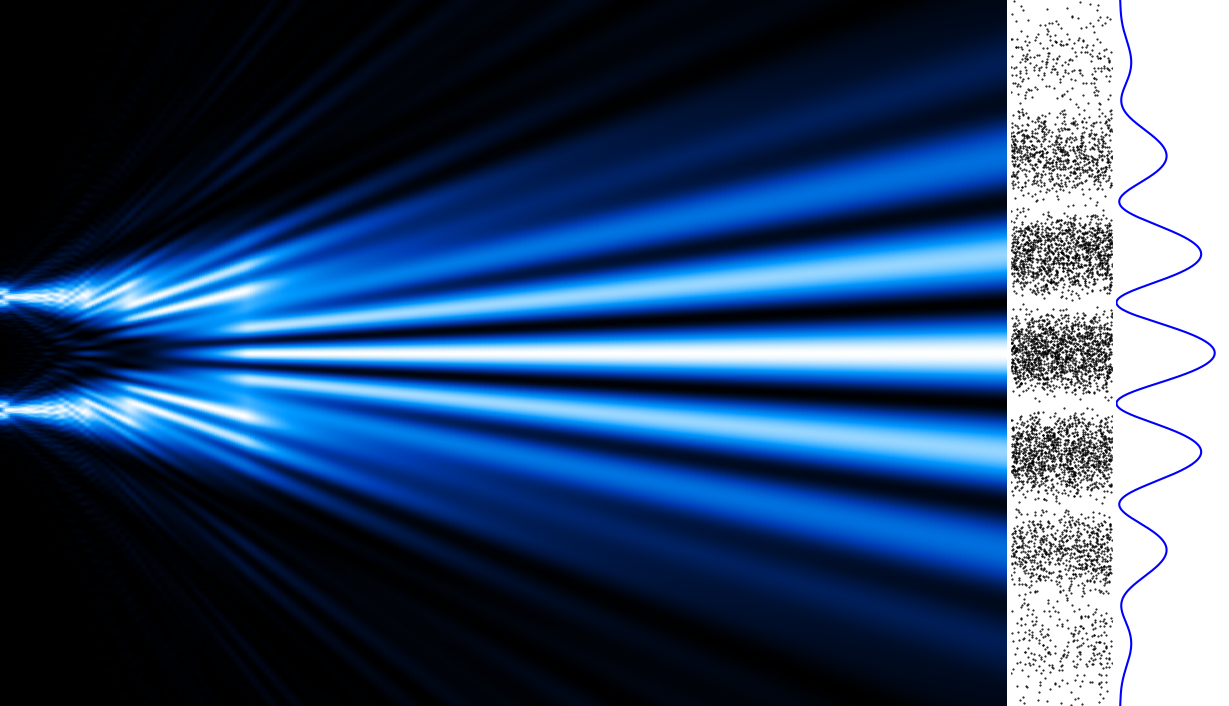}
\caption{\label{fig:results2slits}Simulation results for Hypothesis~2: identical to those from the double-slit experiment. 
\textbf{Left:} The interference pattern (i.e. the norm of the statistical wave function, $|\psi(y)|^2$) for the double-slit experiment alone ($A_1$ and $A_1$ slits), without the $B$ grating on the detection screen at 2m after the slits. This also corresponds to the pattern excepted for the Hypothesis~2 interpretation; it is as if the grating B did not exist. 
\textbf{Right:} Evolution of the probability density $|\psi(x,y,t=x/v_x)|^2$ for the double-slit experiment, from the slits to a distance of 2m beyond them. For $x=2m$ (on the right),  the  density shown in the figure on the left is obtained.}
\end{center}
\end{figure}

According to Hypothesis~2, as $u$-wave is unable to pass through the slits of B grid, then $\psi$-wave does not pass too.
In this interpretation,  the statistical $\psi$-wave function of the Rydberg atoms only passes through the two large slits~$A_1$ and $A_2$ (forty times greater than the physical ``size'').
In this case, whether or not the 40-slit grating is located between the two slits does not change the results of the experiment. We expect to observe a classic double-slit interference pattern as shown in Figures~\ref{fig:results2slits}.
In term of computation, the integration is done only on $A_1\cup A_2$:
\begin{equation}
\psi(y,t)=\int_{A_1\cup A_2}K(y,t\,;\,y_0,t_0)\,\psi(y_0,t_0)\,dy_0.
\end{equation}

\subsection{Hypothesis 3: Interference patterns with a dark band in the center}

Suppose that the Rydberg atoms are emitted one by one from a coherent source. In that case, all the Rydberg atoms have the \emph{same statistical $\psi$-wave}, which propagates through space and splits into several parts after passing through the slits.
According to the double-solution theory, each $u$-wave of Rydberg atom is different because it is located at a different position according to the initial $|\psi_0(y)|^2$ distribution. The $u$-wave is localised in space and has a fixed \emph{size} equal to $0.1\mu m$ in this experiment; its center of gravity, $X(t)$, is the center of gravity of the particle. The $u$-wave does not spread out in space over time, and does not divide into several parts, unless its nature changes (ionisation, chemical reaction, fission, fusion, etc.). Unlike the $\psi$-wave, it can only pass through a single slit. It is the $u$-wave that is measured by the measuring device when it hits the detection screen.  Each $u$-wave follows a well-defined trajectory guided by the $\psi$-wave. If the trajectory of the $u$-wave hits the plate or a slit in the $B$ grating, the Rydberg atom is destroyed (ionization) and these two wave functions also. 

\begin{figure}[ht]
\begin{center}
\includegraphics[width=0.42\linewidth]{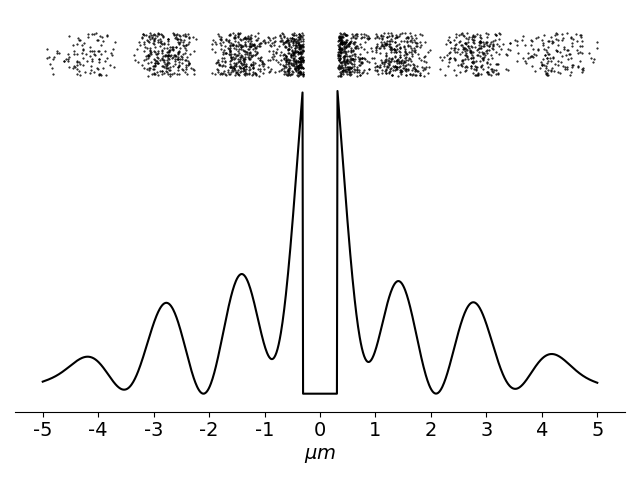}
\includegraphics[width=0.52\linewidth]{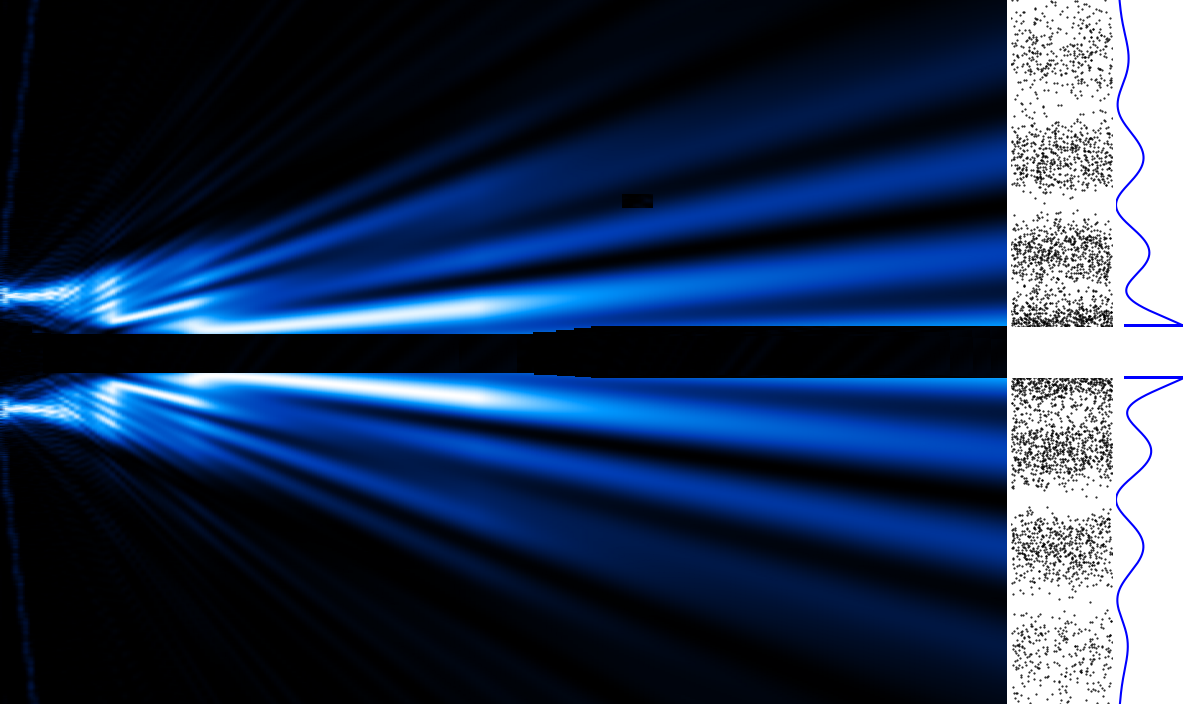}
\caption{\label{fig:results-trucatated}Simulation results for Hypothesis~3: truncated density with a dark band in the center. 
\textbf{Left:} The interference pattern excepted by the Hypothesis~3. Same density as Fig.~\ref{fig:results-unexcited} with a dark band in the center. The central peak is split into two. A Rydberg atom passing through slits $A_1$ or $A_2$ is piloted by the statistical $\psi$-wave function which passes through both silts AND the $B$ grating.
\textbf{Right:} Evolution of the probability density $|\psi(x,y,t=x/v_x)|^2$ excepted by the Hypothesis~3, from the slits to a distance of 2m beyond them. For $x=2m$ (on the right),  the  density shown in the figure on the left is obtained.}
\end{center}
\end{figure}

Only the $u$-waves arriving at one of the two slits $A_1$ and $A_2$ will pass through. The trajectory of the $u$-wave after the slits is guided by its $\psi$-wave function, which is passed through the slits $A_1$ and $A_2$ but also through the 40 slits of the $B$ grating. Its trajectory is therefore different from the one where there would only be double-slit $A_1-A_2$. The density excepted at $2m$ after the slits are shown in Figure~\ref{fig:results-trucatated} and corresponds to that in Figure~\ref{fig:results2slits} truncated by one-third to the center.

The total width of the 40 slits in grid B is equal to the width of slit A1 or A2 ($4\mu m$), that is, one-third of the total width of $A_1\cup A_2\cup B$. It is therefore necessary to truncate one-third of the density $|\psi(y,t)|^2$; since the grid is centered, it is the central trajectories that will not pass through and must be subtracted from the density. Let $Y_{1/3}(t)$ and $Y_{2/3}(t)$ denote the first and second time-dependent tertiles (3-tiles) of $|\psi(y,t)|^2$, respectively~\footnote{For a probability, $p$, Brandt et al.~\cite{Brandt1998} define the \textit{time-dependent quantile position} (quantile trajectory), $Y_{p}(t)$, as: $p=\int_{Y_{p}(t)}^{+\infty}|\psi(y,t)|^2dy$; in our case $p=1/3$ and $p=2/3$.}. The density in Figures~\ref{fig:results-trucatated} is truncated by the band between the two tertiles $Y_{1/3}(t)$ and $Y_{2/3}(t)$.


\begin{equation}
\psi(y,t)=\int_{A_1\cup A_2\cup B}K(y,t\,;\,y_0,t_0)\,\mathbbm{1}_{]-\infty,Y_{1/3}(t)]\cup[Y_{2/3}(t),+\infty[}(y)\,\psi(y_0,t_0)\,dy_0
\end{equation}
with $\mathbbm{1}_{I}(x)=1$ if $x\in I$ and $=0$ otherwise; The density within the interval $[Y_{1/3}(t), Y_{2/3}(t)]$ accounts for one-third of the total density of $\psi$; it must be zero and corresponds to the dark band in the center. 


\section{Discussions and conclusions}

\paragraph{Atomic-metal interaction during the slit crossing.}
The simulations presented in this article do not take into account the atom-metal interactions that occur near the slits; we have made binary approximations that idealize the passage of Rydberg atoms through the slits: either they pass through the slits without being deflected or altered (retaining the same principal number $n$), or they are ionized as they pass through the slits. In reality, the internal structure (i.e. the physical $u$-wave) of the Rydberg atom can change to reach a different $n$ level, or other inelastic atom-metal processes may occur. We must take into account the influence of dispersion related to the width of the slits and the effect of the atom-metal van der Waals attractive force, which curves the atomic trajectories (elastic process~\cite{Parsegian2005}). 
All these corrections modify both the internal $u$-wave and the statistical $\psi$-wave, which may lose its coherence with other branches of $\psi$ and thus be unable to interfere with those other branches. The results will consequently be altered. However, the experimental conditions have been chosen such that these atom-metal interactions are not significant and remain of secondary importance; furthermore, in a real experiment, it will certainly be necessary to re-examine and modify the precise conditions of the experiment. The objective of this paper remains to propose a feasible experiment—even if it is only a thought experiment—rather than to provide a detailed understanding of van der Waals interactions at the level of the slits~\footnote{The discussion of these interactions remains very interesting but also applies to basic double-slit experiments.}. The same type of experiment could also be conducted using large molecules (such as $C_{60}$ or $C_{60}F_{48}$ molecules) and narrow slits.

\paragraph{A better understanding of wave-particle duality}
There is therefore a significant difference between the numerical simulations of the histograms of the impacts of the Rydberg atoms according to the different hypotheses: 
Either Hypothesis 2 is true, in which case we observe a standard double-slit interference patterns (7 peaks; Figures~\ref{fig:results2slits}), and this experiment demonstrates that it exists a \textbf{feedback effect from the $\psi$-wave on the $u$-wave} (i.e. on the particle).
Either Hypothesis 3 is true, in which case we observe a specific interference pattern with a central black band (8 peaks; Figures~\ref{fig:results-trucatated}), and this demonstrates the \textbf{existence of phantom empty waves}.
In both cases, this constitutes a significant advancement within the framework of dBB-type theories with additional variables.


In this article, we focus on the Louis de Broglie’s double-solution theory; however, the results of this experiment have implications for all others interpretations (the Copenhagen interpretation, the Everett interpretation, etc.) that do not assume the existence of a localized physical $u$-wave.
The aim of this article is to show that there is now a feasible thought experiment that can be used to decide between the two points of view, and to deepen our understanding of the problem of wave-particle duality. 

\bibliographystyle{unsrt}
\bibliography{biblio_mq}

\appendix
\section*{Complementary data}
Our code is available on Software Heritage: \url{https://archive.softwareheritage.org/swh:1:dir:16e697c188d37a114248515238b6c44ebde38ddc}\\

\end{document}